\documentclass[a4paper,12pt,reqno,superscriptaddress]{revtex4-1}
\usepackage{graphicx}
\usepackage[centertags]{amsmath}
\usepackage{amsfonts}
\usepackage{amssymb}
\usepackage{amsthm}
\usepackage{newlfont}
\usepackage{stmaryrd}
\usepackage{mathrsfs}
\usepackage{euscript}
\usepackage{color}

\theoremstyle{plain}

\theoremstyle{definition}

\theoremstyle{remark}

\numberwithin{equation}{section}

\newcommand{\bbR}{{\mathbb R}}

\newcommand{\bbZ}{{\mathbb Z}}

\newcommand{\opunit}{\text{1}\kern-0.22em\text{l}}

\DeclareMathAlphabet{\mathpzc}{OT1}{pzc}{m}{it}

\begin{document}

\title{Archimedes' law and its corrections\\ for an active particle in a granular sea}

\author{Christian Maes and Simi R. Thomas}
\affiliation{Instituut voor Theoretische Fysica, K.U.Leuven,
Belgium} %\email{christian.maes@fys.kuleuven.be}
%\author{Simi R. Thomas}
%\affiliation{Instituut voor Theoretische Fysica, K.U.Leuven,
%Belgium}

% --------------------------------------------------------------
\begin{abstract}
We study the origin of buoyancy forces acting on a larger particle moving in a granular medium subject to horizontal shaking and its corrections
before fluidization.  In the fluid limit Archimedes' law is verified; before the limit memory effects counteract buoyancy, as also found experimentally. The origin of the friction is an excluded volume effect between active particles, which we study more exactly for a random walker in a random environment.
The same excluded volume effect is also responsible for the mutual attraction between bodies moving in the granular medium.  Our theoretical modeling proceeds via an asymmetric exclusion process, i.e., via a dissipative lattice gas dynamics simulating the position degrees of freedom of a low density granular sea.
\end{abstract}

\maketitle

% --------------------------------------------------------------------
\section{Introduction}
A main characteristic of granular media is that their behavior varies between being more fluid- and being more solid-like.  Initializing flow via shaking or stirring fluidizes granular baths.  That has been observed in various experiments and simulations, in particular by verifying Archimedes' law, \cite{ex1,ex2,sim1}. Also phenomenological arguments have been added to the understanding of the
buoyancy force in granular media, e.g. from using the Enskog hydrodynamic equations~\cite{phen}. Nevertheless numerous controversies have remained and various corrections must be considered.  In the present paper we take up a simple excluded volume model to study the origin and the corrections to Archimedes' law around the fluid limit.\\

Fluid-like behavior of a granular material should obviously include hydrostatic and hydrodynamic effects.  A natural way to study the origin of granular hydrostatics and their possible corrections is via flow induction, i.e., stirring or shaking the system.  That causes energy transfer to the grains which is dissipated again in the collisions between the moving grains.  Further simplifications can help to understand the essence of what happens.  In that spirit we consider
the asymmetric exclusion process to simulate the dynamics of the grains (monomers) with one large particle (rod) immersed in them.  The condition of detailed balance enables us to identify the buoyancy force on the rod, as function of its size and of its relative weight, and locally as function of the height.We recover Archimedes' law in the fluid limit, and we create a theoretical framework for a detailed study of possible corrections. Corrections
arise from various effects such as from the discrete nature of the lattice where the lattice spacing measures the size of the grains. Corrections also arise from thermal effects especially when the grains are themselves immersed in a heat bath (e.g. hot air), and as studied here, from finite shaking rates. Other possible corrections arise from convection currents in the granular medium, to which we turn briefly in Section \ref{rema}.\\

To the extent that our modeling via the asymmetric exclusion process is relevant for the experimental conditions of granular media under shaking, Archimedes-like behavior was already predicted in \cite{hydro}. Five years later Archimedes' law was confirmed experimentally \cite{ex1} in a granular medium, and some corrections were explored before the fluid limit.
In the mean time further experimental work and simulations which show an explicit Archimedes like behavior have been added, including \cite{ex2,sim1}. We come back to the basic set-up.\\

\subsection{Phenomenon \& Experiments}

The experiment \cite{ex1} consists of a bi-dispersion of glass beads which form the granular bath. This bath is placed in a rectangular box, on which horizontal shaking is applied through vibrations on the walls.
Gravity controls the vertical motion. The bath fluidizes for larger energy input through the external shaking.
The energy of shaking is quantified by a dimensionless quantity $\Gamma \propto f^2 A/g$ related to frequency $f$ and amplitude $A$ of
shaking which in the case of horizontal shaking exercise a similar effect.
One observes a clear transition; before some energy threshold is reached the state is mostly solid  and there is a clear
boundary where the unfluid state becomes fluidized and the principle of Archimedes is obeyed henceforth.
The correction to Archimedes' law can be described by a drag force.  Several experiments have studied the behavior of this drag force \cite{ex1,ex2,Mob,creep,slowdrag}. Most agree that in the fluidized and low density limit the drag force is
observed to be proportional to the velocity. Moreover the coefficient of viscosity is seen to vary exponentially with the shaking amplitude \cite{ex1,Mob}, or, in \cite{ex2}, observed inversely proportional to the shaking amplitude. For denser media the drag force was seen to be varying logarithmic with velocity \cite{creep}.
In our model we discuss the nature of drag force and corrections to it in Section \ref{gmasmall}.

\subsection{Model and results}

The goal of this paper is to give a simple theoretical model giving a natural interpretation of these experimental results, and at the same time to allow further studies on modifications of Archimedes' principle under
the influence of fluctuations and nonequilibrium effects.
On a more fundamental level, our model illustrates the construction of statistical forces.
We have here an example of how the effective motion of the bigger particle gets changed indirectly by the influence of that same particle on the smaller particles. In other words, the rod (bigger particle) is active in changing its environment.  The result can be characterized as an additional effect modifying Archimedes' buoyancy before the fluid limit.
  %The models and our results are therefore very compatible with the findings in \cite{,monas} concerning effective motion of active particles, %except that here we apply these to correct Archimedes' law as experimentally studied in \cite{ex1}.\\

The granular medium that is considered are particle conserving lattice gases that evolve under excluded volume conditions.  The configuration is completely determined by the particle occupations (there are no velocities) but the dynamics is dissipative and therefore simulates to some extent the behavior of low density granular materials.\\
The plan of the paper is as follows. In the next section we describe our model, introduced in \cite{hydro} which is an exclusion process of walkers on a regular lattice in the presence of a spatially extended object. In Section \ref{fluidlimit} the fluid limit yields Archimedes' law in terms of a conservative buoyancy force proportional to the local fluid density times the volume of the object.
Further the continuum limit to the spatial degree of freedom reveals the first level of corrections to the nature of buoyancy force. Fluctuations due to the surrounding heat bath bring in the next level of correction to the usual hydrostatic formulation.\\
The main result appears in Section \ref{redgreen} where we investigate corrections to the fluid limit.  We map our model to a simpler one-dimensional problem, to obtain the motion of a random walker in a stochastic dynamical environment.  There, the correction to Archimedes' law is explained by a memory effect which acts as friction on a rising object. Continuum corrections bring additional clarification.
We discuss how our model adds a simple heuristic picture to the experimental observations of \cite{ex1}.
The last sections discuss additional aspects that become visible through our theoretical modeling. A discussion on the existing mechanisms of intruder dynamics in granular media and a comparison with our model appears in Section \ref{rema}. In section \ref{rema1} results of simulations modeling our system are shown. %, like the construction of statistical forces %outside equilibrium.\\
Future possibilities of nonequilibrium effects and collective behavior are alluded upon in the last sections.

\section{A rod in a lattice fluid}\label{fluid}

We consider a
stochastic dynamics for the motion of a rod in a
lattice fluid composed of monomers. The model was introduced in \cite{hydro}, obtaining a Markovian reduced
dynamics for the rod motion in the limit where between any two moves of the rod, the monomer
configuration has the time to relax.  That can be interpreted as the fluid limit in which the rod always ``sees'' the fluid of monomers in equilibrium.

\subsection{Model}

\begin{figure}[ht!]
 \begin{center}
\includegraphics[height=5cm]{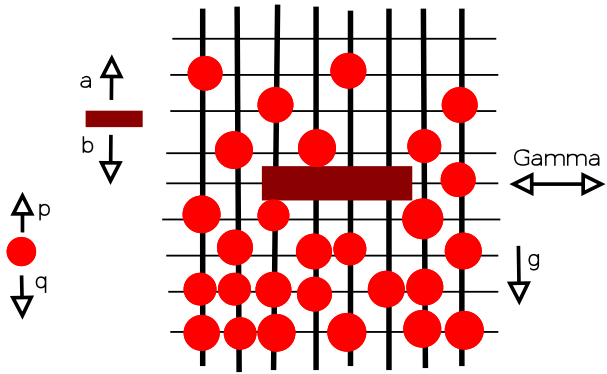}%,width=5in
\caption{Two-dim model}
\end{center}
%\caption{}
\end{figure}

All motion takes place on the square lattice $\bbZ^2$ where the mesh size, taken unity here, gives the size of the grains.  These grains (also called, monomers) can occupy sites $i=(x,y)$ having ``vertical'' coordinate $y$ and
``horizontal'' coordinate $x$. There can be at most one grain per site.  There is one big particle or big grain, called rod and we only follow its vertical position.  The vertical position of the rod
at time $t$ is denoted by $Y_t$ taking values in $\bbZ$.  The rod occupies $N\in\{ 2,3,\ldots\}$ lattice sites in the horizontal direction, i.e., the region
\begin{equation}\label{regi}
A_N(y)= \{(0,y),(1,y),\ldots,(N-1,y)\}
\end{equation}
 is forbidden for the
monomers when  $Y_t=y$.\\
The
horizontal jumps of the monomers are symmetric at
rate  $\gamma$. Increasing the rate $\gamma$  speeds up the monomer dynamics in the horizontal direction (orthogonal to the motion of the
rod).  The vertical jumps of the monomers and the rod are asymmetric; modeling a gravitational field.
 Note that the vertical motion is not speeded up, but of course it is influenced by $\gamma$ as well.\\

More formally, the microscopic dynamics looks as follows; see Fig.~1:
 The monomer configuration is denoted
by $\eta\in \{0,1\}^{\bbZ^2}$; $\eta_t(i) = 0$
means there is no monomer at site $i$ at time $t$ and $\eta_t(i) = 1$ if
there is a monomer at site $i$ at time $t$.  The dynamics  is of exclusion--type because all motion is via jumping to vacant sites, and rod and monomers never overlap.
A monomer moves horizontally to a vacant nearest neighbor site, symmetrically with rate $\gamma$.  It moves vertically up with rate $p$ and down with rate $q$.  The rod only moves vertically, up with rate $a$ and down with rate $b$.
We choose $p/q,a/b < 1$ to represent the gravitational field, e.g. via
$p/q = \exp (-mg/kT), a/b= \exp (-Mg/kT)$, where
$M, m$ denote the mass of the rod, respectively of the monomers; $kT$ is a typical unit of thermal energy at temperature $T$ which plays little role in what follows, except for allowing fluctuations. The temperature could also refer to an additional heat bath that makes contact with the grains. Yet, granular media are typically a-thermal in which case we think of $Mg,mg >> kT$.  The lattice unit is not indicated; it is taken to be one and should be thought of as the size of the grains.   All that motion gets summarized in the formal generator
\begin{eqnarray}\label{lori}
Lf(\eta,y) \equiv && a\,I[\eta(i)=0, \forall i\in A_N(y+1)]
[f(\eta,y+1)
- f(\eta,y)]\nonumber\\
& +&\; b\, I[\eta(i)=0, \forall i \in A_N(y-1)]\,
[f(\eta,y-1)
- f(\eta,y)] + \nonumber\\
&& \sum_{i=(i_1,i_2)} \Bigl\{p\, \eta(i)\, (1-\eta(i_1,i_2+1))\, I[(i_1,i_2 + 1) \notin
A_N(y)]\,\times[f(\eta^{i,(i_1,i_2+1)},y) - f(\eta,y)]
\nonumber\\
&+& q\, \eta(i)\, (1-\eta(i_1,i_2-1))\, I[(i_1,i_2-1)
\notin A_N(y)]\times[f(\eta^{i,(i_1,i_2-1)},y) - f(\eta,y)]\Bigr\} \nonumber\\
&+&
\gamma
\sum_{\langle ij\rangle:i_2=j_2}
I[\langle ij \rangle \cap A_N(y) = \emptyset]
[f(\eta^{i,j},y)
- f(\eta,y)]
\end{eqnarray}
where $I[\cdot]$ is the indicator function of the event in the brackets, giving one or zero depending on the event being realized, and $\eta^{i,j}$ is the grain configuration $\eta$ after switching the occupations in sites $i$ and $j$.  The last term represents the horizontal shaking in which the occupations of (horizontal) nearest neighbor pairs $\langle ij \rangle$ get exchanged.  Observe that there is always both horizontal and vertical motion, subject to the exclusion rule, which, besides from the shaking, can arise from an extra heat bath in which the grains are moving and with which energy can be exchanged.\\
 We need \eqref{lori} for writing down the kinetic equations that are all of the form
\[
\frac{d}{dt}\langle f(\eta_t,Y_t) \rangle = \langle Lf\, (\eta_t,Y_t) \rangle
\]
Here and from now on, brackets $\langle \cdot\rangle$ are with respect to the stochastic dynamics and over the following initial conditions.
At starting time $t=0$ we put
$Y_{t=0}=0$ so that the rod starts from the center of the lattice, but that is really arbitrary.
 For the initial distribution on monomer occupations we take density
\begin{equation}
\label{dens}
d(x,y) = d(y) = \frac{\kappa (p/q)^{y}}{1 + \kappa (p/q)^{y}}
\end{equation}
for parameter $\kappa >0$. This formula satisfies $p\,d(y)\,(1-d(y+1))=q\,d(y+1)\,(1-d(y))$ which is a detailed balance relation for the motion of the grains, see also \eqref{dba} below.  The density varies between zero (at the top) to one (at the bottom). The height where $d(y)=1/2$ scales like $y \sim \log \kappa$.  The derivative of the density at that height (where the transition is made between higher and lower density) is proportional to $mg/kT$.  Therefore, choosing $\kappa\simeq 1$ and large $mg/kT$ corresponds to a more constant density as in a liquid or as in a granular medium in a container filled from the bottom to around $y=0$;  on the other hand, looking at positive $y$, for $\kappa=1$ and for smaller $mg/kT$ corresponds to a gas condition where \eqref{dens} simulates a barometric formula. Low density granular media under heavy shaking would also fall in that category which in our modeling is most typical.    The density is constant in the horizontal
direction ($x$), but always conditioned on having $\eta(i)=0$ for all $i$ covered by the rod, i.e., for all
$i \in A_N(Y)$.\\
More precisely, we let $\nu_{d}$ denote the product measure on
$\{0,1\}^{\bbZ^2}$ with density
\begin{equation}\label{den}
\text{Prob}[\eta(x,y) =1] = d(y)
\end{equation}
defined by (\ref{dens}). The conditional probability
\begin{equation}
\nu_d^0 = \nu_d( \cdot | \eta(i)=0, \forall i \in A_N(0))
\end{equation}
is then the initial distribution on the monomers.  The dynamics such as defined
above gives rise to the Markov process $(\eta_t,Y_t)$.\\

\subsection{Fluid limit} \label{fluidlimit}
In the limit $\gamma\uparrow +\infty$ the motion of the rod
decouples from the
monomer dynamics. Then, the reduced
dynamics of the rod becomes that of a random walker with rates
directly given in terms of the equilibrium fluid density:\\
the rod moves up $y\rightarrow y+1$ with new rate $a[1-d(y+1)]^N$ and goes down $y\rightarrow y-1$
with new rate $b[1-d(y-1)]^N$.  The factors $[1-d(y\pm 1)]^N$ of course express the plausibility
 of having space for the rod to move from height $y$ to $y\pm 1$; there must be a hole of size $N$.
Hence, in the limit of excessive horizontal shaking, the rod is doing a
continuous time random walk on $\bbZ$ with backward generator
\begin{equation}\label{rw}
L^{\textrm RW}f(y) = a[1-d(y+1)]^N[f(y+1)-f(y)]+
b[1-d(y-1)]^N[f(y-1)-f(y)],
\end{equation}
The density profile $d$ is obtained from (\ref{dens}).\\

We call this limit $\gamma \uparrow \infty$ the fluid limit.  The reason of the decoupling is that the monomers relax to their stationary reversible density in between any two moves of the rod. The resulting motion \eqref{rw} is itself
satisfying the condition of detailed balance for a potential $V$, which can be interpreted as giving rise to a conservative force $F$ given by the logarithmic ratio of up {\it versus} down rates
\begin{eqnarray}\label{pot}
F(y) = -V(y) + V(y-1) &=& -\,kT\,\ln \frac{b[1-d(y-1)]^N}{a[1-d(y)]^N}\nonumber\\
&=&  -Mg - NkT\,\ln [1 + \frac{d(y)-d(y-1)}{1-d(y)}]\nonumber\\
\end{eqnarray}
To go to a continuum description (at least in the vertical direction) we introduce a lattice mesh of size $\varepsilon>0$, under which $\ln a/b = -Mg\varepsilon/kT, \ln p/q = -mg\varepsilon/kT$.  This imagines that the grains are of vertical size $\varepsilon$.  We compute the force $F_\varepsilon$ as $\varepsilon\downarrow 0$:
\begin{equation}\label{pote}
F_\varepsilon(y) = \frac{ -V(y) + V(y-\varepsilon)}{\varepsilon} =-Mg - \frac{kT\,N}{\varepsilon}\,\ln [1 + \varepsilon\frac{d'(y)}{1-d(y)}]
\end{equation}
where the density $d(y)$ is now on $\bbR$ and, similar to \eqref{dens}, verifies
\begin{equation}\label{derden}
d'(y) = -d(y) (1-d(y)) \frac{mg}{kT}
\end{equation}
Hence, \eqref{pote} becomes $F_\varepsilon \rightarrow F$ for
\begin{equation}\label{arc}
F(y) = -Mg + d(y)\,mg\,N
\end{equation}
which is Archimedes' law for the total upward force on a body of volume $N$ and mass $M$ replacing a weight equal to $mgd(y)N$ of fluid.  For short, we call this the buoyancy force. The rod will thus move to a height
where the fluid density is proportional to $1/N$.  That is the equilibrium position, consistent with Archimedes' characterization of the hydrostatic equilibrium position, \cite{Archi}.  The force \eqref{pot} is the correction
to the Archimedes' force \eqref{arc}, due to the finite size of the grains.\\

The motion can be studied in the diffusive limit where we also rescale time $\sim \varepsilon^2$.  That means to take for example
\begin{equation}\label{cho}
a = \frac{e^{-Mg\varepsilon/(2kT)}}{\varepsilon^2},\quad b = \frac{e^{Mg\varepsilon/(2kT)}}{\varepsilon^2}
\end{equation}
and to expand the generator
\[
L_\varepsilon^{\textrm RW} f(y) = a\,[1-d(y+\epsilon)]^N\, [f(y+\epsilon) -f(y)] +
b\,[1-d(y-\epsilon)]^N \,[f(y-\epsilon) -f(y)]
\]
(see \eqref{rw}) in orders of $\varepsilon$.  The result is that $L_\varepsilon^{\textrm RW} f(y) \rightarrow Lf(y)$
with
\begin{equation}\label{overd}
Lf (y) = (D\,\,f')'(y) + \chi(y)\,F(y)\,f'(y)
\end{equation}
with
\[
F(y) = Nd(y) mg - Mg, \;\;\; D(y) = [1-d(y)]^N,\;\;\chi(y) = \frac{D(y)}{kT}
\]
We again have made use of \eqref{derden} in the continuum limit.
The result \eqref{overd}  is the generator of an overdamped diffusion equation with diffusion coefficient $D(y)$.
The corresponding Langevin equation, in the It\^o-sense, is given by
\begin{equation}\label{lange}
\dot y_t = \chi(y_t)\,F(y_t) + D'(y_t) + \sqrt{2D(y_t)}\,\xi_t
\end{equation}
for white noise $\xi_t$.
The force $F$ is exactly the one found in \eqref{arc}, as in Archimedes' law.
The diffusion $D(y)$ is related to the mobility $\chi(y)$ via the Einstein equation $kT\,\chi(y) = D(y)$.
The term with the derivative $D'(y)$ is due to the It\^o-convention.\\

The above analysis concludes that the two--dimensional lattice model on a lattice with a simple exclusion dynamics, provides a reasonable description of the hydrostatic behavior of granular matter in the
limiting (fluidized) case. We have seen above how the discreteness of the vertical lattice-direction makes a first correction, easily studied for small lattice mesh.  A second type of correction is due to fluctuations.
The rod dynamics generated by \eqref{rw} is stochastic.  The fluctuations of the rod about its equilibrium position can be studied in the large $N$ limit with
standard deviation around the mean going as $1/\sqrt{N}$.
For the third major type of correction, we study the approach to the fluid limit.
The next section is devoted to these questions.

\section{Random walk in a dynamical environment}\label{redgreen}

\begin{figure}[ht!]
\begin{center}
\includegraphics[height=5cm]{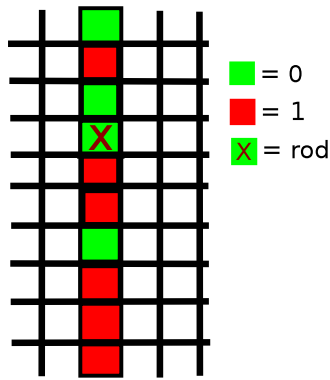}
\caption{Contracted description}
\end{center}
%\caption{Fig.2: }
\end{figure}

To this end we propose a contracted description of the model, coarse-graining it to an effective one-dimensional model; see Fig.~2. The idea is as follows.  The essential aspect of the monomer dynamics as far as regards the rod, is whether there is a hole above or below the rod in which it can jump.  We therefore summarize all of the monomer configuration $\eta(x,y)$ by variables $\sigma(y), y\in \bbZ$, that specify whether or not there is a monomer in the region $A_N(y)$, see \eqref{regi}:
 \begin{eqnarray}
\sigma(y) &=& 1 \text{  if there is no hole at } y\\
&=& 0 \text{    if there is a hole at } y
\end{eqnarray}
More precisely, there is a hole at $y$ if $\eta(i)=0$ for all $i\in A_N(y)$.
The position of the rod is still denoted by $Y$.  We assume as major simplification that the (contracted) system $(\sigma_t(y), Y_t), y\in \bbZ, t\geq 0$, undergoes a joint Markov process which mimics the original model in the following sense.\\
The rod moves up $y\rightarrow y+1$ with rate $a$ if there is a hole at $y+1$, i.e., if $\sigma(y+1)=0$.
The rod moves down $y\rightarrow y-1$ with rate $b$ if there is a hole at $y-1$, i.e., if $\sigma(y-1)=0$.
The rod never moves to a position $y$ where there is no hole, $\sigma(y)=1$.  For the monomer dynamics, we assume that the $\sigma_t(y)$  flip $0 \rightleftarrows 1$ with different rates depending on $y$, and depending on the position of the rod.  More precisely, $\sigma_t(y)$ has rate $q(y)$ for the change $1\rightarrow 0$ and has rates $p(y)$ for $0\rightarrow 1$ except when $Y_t=y$ because then it must remain zero; see Fig.~3.
\begin{figure}[ht!]
\begin{center}
\includegraphics[height=5cm]{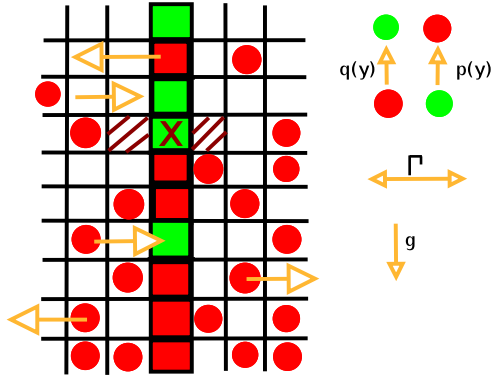}
\end{center}
\caption{Contracted dynamics}

%\caption{Fig.2: }
\end{figure}

So formally the backward generator of our new Markov process is
\begin{eqnarray}\label{coa}
\nonumber
{\cal L}f(\sigma,Y)
&=& a[1-\sigma(y+1)][f(\sigma,y+1) - f(\sigma,y)]\\
\nonumber
&+& b[1-\sigma(y-1)][f(\sigma,y-1)-f(\sigma,y)]\\
&+& \sum_{y\in Z}\big[[1-\sigma(y)]\,p(y)(1-\delta_{y,Y}) + \sigma(y)\,q(y)\big]\,[f(\sigma^y,Y) - f(\sigma,Y)]
\end{eqnarray}
where $\sigma^y$ is the hole-configuration obtained after flipping the occupation at $y$: $\sigma^y(y') =\sigma(y')$ if $y\neq y'$ and $\sigma^y(y') =1-\sigma(y)$ if $y=y'$.   At no point in time could the region occupied by the rod be simultaneously occupied by a monomer and vice versa,
so $\sigma_t(Y_t) = 0$ always.  This is the influence of the active particle (the rod) on the fast degrees of freedom.

\subsection{Interpretation}
Obviously, the rates $p(y), q(y)$ must be interpreted in terms of the monomer density $d(y)$ at $y$ with their dependence on the size $N$ of the rod and on the amount of horizontal shaking $\gamma$.  Comparing \eqref{coa} with
\eqref{lori} suggests further interpretations.\\

It remains that
 $a/b = \exp(-Mg/k_BT)$  where $M$ is the mass of the rod.
We can think of the monomers as blinking lights, red ($\sigma(y)=1$) for no passage of the rod, and green ($\sigma(y)=0$) for passage allowed.  In the original two-dimensional model a hole (green light) $\sigma(y)=0$ at $y$
represents the fact that there are no monomers at the sites $A_N(y)= \{(0,y),(1,y),\ldots,(N-1,y)\}$, and (red light) $\sigma(y)=1$ means that some monomer can be found in the region $A_N(y)$. Abbreviating
\[
\rho(y) \equiv 1 -  (1 - d(y))^N
\]
we take therefore
 \begin{equation}\label{cho}
 p(y) = \gamma\,\rho(y),\quad
 q(y) = \gamma\,(1-\rho(y))
 \end{equation}
for the rate at which a hole gets removed, respectively created. Each depend on $N$ and on $y$ but observe that  $p(y) + q(y) = \gamma$ which is the horizontal shaking rate.   The stationary hole density at $y$ for that two-state Markov process becomes
$1- \rho(y) = (1-d(y))^N$ which is the correct hole probability in the original monomer-model, cf. \eqref{den}.  Of course the weight of the monomer is represented in the density $d(y)$ via \eqref{dens}.\\

Here also, for our simplified model we can take the fluid limit $\gamma \uparrow \infty$.  By simpler arguments than in \cite{hydro}, the motion of $Y_t$ decouples from that of the $\sigma_t$ and by our choice \eqref{cho} we find exactly the same limiting motion of the rod as given in \eqref{rw}.  The two models mathematically agree in the fluid model but our second model allows more easily to find the most significant contribution before the fluid limit, to which we turn next.

\subsection{Before the fluid limit}\label{gmasmall}

Suppose we find the rod at time $u$ in position $y$.  Thus, at that time the hole probability at $y$ is equal to 1.
 When the rod jumps from position $y$ to say position $y+1$ at time $u$ it leaves a ``hole'' at position $y$ which remains a hole until it gets occupied by either monomers or
by the rod again.
In the fluid limit the monomer dynamics is fast enough and they relax to their equilibrium configuration so that at its next jump the rod sees a hole with probability $1-\rho(y)$. On the other hand if the
time of monomer relaxation is longer in comparison to the rod, at the next jump  at time $u+t$ the rod sees ``no hole'' with probability
\begin{equation}\label{noh}
 \rho(y)(1-\exp(-\gamma t))
\end{equation}
as follows from a simple calculation for the two-state Markov process $\sigma_t(y)$.  Note however that the transient density \eqref{noh} is lower than $\leq \rho(y)$ and that \eqref{noh} is only valid under the condition that the rod has not (re-)entered position $y$ during $[u,u+t]$.  It implies that at time $u+t$ the rod still sees the hole it left behind at $y$. We conclude that before the fluid limit, the jump rates of the rod depend also on the past and the rod dynamics is by itself non-Markovian for finite $\gamma$.\\

We make the above statements now more concrete, and more precise.
Knowing the rod's motion follows from the evolution equation, for all functions $g$ on $\bbZ$,
\[
\frac{d}{dt}\left<g(Y_t)\right> = \left<{\cal L}g\,(\sigma_t,Y_t)\right>
\]
with, from \eqref{coa},
\begin{equation}
{\cal L}g\,(\sigma,y) =  a[1-\sigma(y+1)][g(y+1) - g(y)]
+ b[1-\sigma(y-1)][g(y-1)-g(y)]
\end{equation}
Consider for example the expectation
\[
\left<[1-\sigma_t(Y_t+1)]g(Y_t+1)\right>=
 \langle\left<1-\sigma_t(Y_t+1)|Y_s, 0\leq s \leq t \right>\,g(Y_t+1)\rangle
 \]
  where
  \begin{equation}\label{ho}
  \left<\sigma_t(Y_t+1)|Y_s, 0\leq s \leq t \right>\end{equation}
   is the
conditional probability of having no hole just above the rod,  given the full history of the walker $(Y_s, 0\leq s \leq t) $.
Obviously, history matters.\\
Suppose for example that at time $t$ we have $Y_t=y$ and that the rod has been there already for a time $t_1$.  The previous position was either $y+1$ or $y-1$, from which the rod has moved at time $t- t_1$.
Before that, at jump time $t-t_1 - t_2$ the rod has been jumping either from $y-2$, from $y$ or from $y+2$, {\it et cetera}.  In this way the whole history of the rod can be parameterized in terms of waiting times and successive positions.  We denote such a rod-history by $\omega$. Yet, the only thing that matters for the expected hole probability at $y+1$ in \eqref{ho} is the last time $t(\omega,y+1)$  it was occupied by the rod, since
\begin{equation}
 \left<[1-\sigma_t(Y_t+1)]\,|\,\omega\right> = (1- d(y+1))^N + \rho(y+1)\,\exp(-\gamma (t-t(\omega,y+1)))
\end{equation}
We put $t(\omega,y)=-\infty$ if the rod has never been in position $y$, to realize the initial condition \eqref{den}.\\

We must now estimate the conditional expectation
\begin{equation}\label{cex}
\langle e^{-\gamma[t-t(\omega,y+1)]}|(Y_s,s\in [0,t])=\omega\rangle
\end{equation}
for a history $\omega$ in which $Y_t=y$.  Clearly, that equals $\exp - \gamma t_1$ if before $y$ the rod was at $y+1$; otherwise (if before the rod was at $y-1$)  \eqref{cex} is certainly less than $\exp -\gamma(t_1+ t_2)$, which is much smaller than $e^{- \gamma t_1}$ for large $\gamma$.  There are then two cases depending on the sign of the rod's ``velocity''
\begin{equation}\label{rove}
V_t= Y_t - Y_{t-t_1}
\end{equation}
We therefore approach the fluid limit by putting \eqref{cex} equal to zero if  $V_t$ is positive, and
by putting it equal to
\begin{equation}\label{fri}
\mu(\gamma)\equiv \int_0^{+\infty} e^{- \gamma t_1} \,(a+b)\,e^{-(a+b)t_1}\,dt_1 = \frac{a+b}{a+b+\gamma}
\end{equation}
if $V_t$ is negative. The integral \eqref{fri} takes   the expectation over the exponential waiting time distribution for $t_1$.  In \eqref{fri},
the sum $a+b = v$ is a good estimate for the average speed of the rod, or $\mu(\gamma)\,V_t =  \frac{vV_t}{v+\gamma} = \nu(\gamma)\,\vec{v}_t$, where $\nu(\gamma) = 1/(v+\gamma)$ is the friction coefficient and $\vec{v}_t$ is the velocity of the rod before it arrived at $Y_t$.\\
The drag force on a particle immersed in granular matter was studied in various experiments --- in \cite{ex1,ex2,Mob,creep} a linear dependence on the particle velocity such as proven above was observed and corresponds to low density.
Inserting this $\mu(\gamma)$ we have obtained for large but finite $\gamma$ that
\begin{eqnarray}\label{rwb}
 \nonumber
 \frac{d}{dt}\left<g(Y_t)\right>
&=& \left<L^{\textrm RW}g(Y_t)\right>\\
\nonumber
&+& a\left<\rho(Y_t+1)\mu(\gamma)\frac{[1-V_t]}{2}[g(Y_t+1)-g(Y_t)]\right>\\
&+& b\left<\rho(Y_t-1)\mu(\gamma)\frac{[1+V_t]}{2}[g(Y_t-1)-g(Y_t)]\right>
\end{eqnarray}
always with $\rho(y)= 1 - [1-d(y)]^N$ and the first line of \eqref{rwb} corresponds to the fluid limit \eqref{rw}.  The $V_t$ is $\pm 1$ as defined in \eqref{rove}.\\

To recapitulate, the approximation in which we replace \eqref{cex} by \eqref{fri} is the following. The rate of the rod's dynamics in comparison to the monomer dynamics is such that for the rod making a jump $y\rightarrow y+1$  it can still ``see'' the gap it left at $y+1$ when indeed the rod was at $y+1$ before it came to $y$.  However the rod does not ``see'' any gaps which were left at $y+1$ from earlier visits there:  the monomer dynamics is fast so that it can erase the trace of the rod's
 trajectory up to one time step ago.  The introduction of the velocity $V_t$ of the rod is a way to re-install the Markov property, where the state of the rod is now defined as its position plus its (previous) velocity.  In other words, due to the active nature of the rod it acquires memory before the fluid limit (which results in a drag force, see below), which is most efficiently dealt with by introducing a velocity.\\

For the position of the rod, $g(y) = y$,
\begin{eqnarray}
 \nonumber
 \frac{d}{dt}\left<Y_t\right> &=& \left<a(1-\rho(Y_t+1)) - b(1-\rho(Y_t-1))\right>\\
\nonumber
&+& a\left<\rho(Y_t+1)\mu(\gamma)\frac{[1-V_t]}{2}\right>
- b\left<\rho(Y_t-1)\mu(\gamma)\frac{[1+V_t]}{2}\right>
\end{eqnarray}
This equation gives the speed of the rod at time $t$ given its current position and previous direction $V_t$. If the rod was moving upwards ($V_t= +1$), then it continues moving up
with a rate $a\left<(1-\rho(Y_t+1))\right>$ and goes down with a rate $\left<b(1-\rho(Y_t-1))\right> + b\,\mu(\gamma)\,\left<\rho(Y_t-1)\right>$. On the other hand if the rod was moving downwards, $V_t= -1$, then it continues moving down
with a rate $b\left<(1-\rho(Y_t-1))\right>$ and goes up with a rate $\left<a(1-\rho(Y_t+1))\right> + a\,\mu(\gamma)\,\left<\rho(Y_t+1)\right>$. In comparison to the fluid-limit there is an increase in the rate
of return. The rod has a higher tendency to go back to the site it started from when the bath is not completely fluid. The rate to go forward remains the same as in the fluid limit. This phenomenon
of a greater tendency to return  with possible subsequent oscillations can be interpreted as a greater dynamical activity which becomes effective as the bath becomes less and less fluid.  When the rod has an overall tendency of rising because of the greater buoyancy, the result is friction acting downwards.\\

As for the fluid limit here also we can make  a  small mesh analysis and take the diffusive limit.  We also need to rescale the shaking $\gamma \rightarrow \gamma/\varepsilon^2$ so that with the choice of \eqref{cho}, $\mu(\gamma) \rightarrow 2/(2+\gamma)$ and $V_t = \pm\epsilon$.
We only need to worry about the additional last two lines in \eqref{rwb}, i.e., corresponding to
\begin{eqnarray}
 \nonumber
&& \varepsilon^{-2}\,(1 - \frac{Mg\varepsilon}{2kT})\left[1-{(1-d(y+\varepsilon))}^N\right]
\frac{\varepsilon^-}{2+\gamma}[g(y+\varepsilon) - g(y)]\\
&+& \varepsilon^{-2}\,(1 + \frac{Mg\varepsilon}{2kT})\left[1-{(1-d(y-\varepsilon))}^N\right]
\frac{\varepsilon^+}{2+\gamma}[g(y-\varepsilon) - g(y)]
\end{eqnarray}
where $\varepsilon^- = 2\varepsilon$ if the rod was going down and $\varepsilon^-=0$ when the rod was going up; similarly
$\varepsilon^+ = 2\varepsilon$ if the rod was going up and $\varepsilon^+=0$ when the rod was going down.
Again making the $\varepsilon-$expansion we find, similar to \eqref{overd}, the corrected
Langevin equation
\begin{equation}\label{lange1}
\dot{y_t} = \chi(y_t)\,F(y_t) - \frac{2(1-D(y_t))}{2+\gamma} \upsilon_t + D'(y_t) + \sqrt{2 D(y_t)}\,\xi_t
\end{equation}
with memory term in the friction, $\upsilon_t =  \frac{y_t - y_{t-dt}}{\Arrowvert y_t - y_{t-dt}\Arrowvert} $ being the direction of the velocity just before time $t$;
the rest of the Langevin equation \eqref{lange1} is interpreted in the It\^o-sense with, in particular the left-hand side referring to $y_{t+dt} - y_t$. The diffusion coefficient remains the same as before in the fluid limit, see \eqref{lange}.  The friction $\sim(1-D(y))$ increases with higher density.\\
%\begin{figure}[ht!]
 %\begin{center}
%\includegraphics[height=5cm]{gmvsforce.eps}%,width=5in
%\caption{Buoyancy force versus $\gamma$}
%\caption{Buoyancy force {\it vs} $\gamma$}
%\end{center}
%\caption{Fig.4: }
%\end{figure}

If we look at the origins of drag or friction in common phenomena like Brownian motion, it arises due to a resistance to motion in the form of collisions from the front. The faster a tracer particle moves in a thermal bath
the more traffic it finds ahead of itself than behind.  Of course friction appears in all directions against motion and exists at shaking of all strengths. The drag force we see here is a variation of this effect. Our system is overdamped and nothing of impact or momentum transfer can be discussed; yet interaction via excluded volume will be sufficient to generate (another) force which opposes motion of the rod. This force appears
when granular baths are not completely fluidized.

The higher the intruder dynamics rate, the stronger its memory of its previous position and the greater is the chance of jumping to its old position. The force become weaker and weaker in the fluid limit since the memory of the rod is ``instantaneously'' being wiped away by the monomers.
We believe that the ``drag force'' dependence on $\gamma$ and intruder velocity  as seen in \cite{ex1,ex2} away from the fluid limit are explained by this new kind of opposing force rather than the conventional understanding of friction in fluids, especially in a low density environment where momentum transfer does not play such a big role.

\section{Further remarks}\label{rema}

\subsection{Segregation effect} Effects of buoyancy in granular media have been widely studied both theoretically and experimentally.  It is not always easy to distinguish between anti-gravity effects and buoyancy as in Archimedes' law.  That connects with the variety of segregation effects in granular media upon shaking them.  Reference \cite{Vib} discusses various mechanisms that can work together for the segregation of grains.\\
Buoyancy, i.e., rising/sinking due to pressure gradients, dominates when the fluidization occurs with no convection.  Our model does not show boundary effects and inertia is absent.  The limiting motion is overdamped in the diffusion limit.\\
Buoyancy is indeed most visible in a vibro-fluidized regime, where only binary collisions are prevalent and there is no long time contact between particles. The medium must have minimal convection, so the boundaries must be far and the interactions with the boundary reduced. In a fluidized regime the effects due to convection as well as inertia are reduced enough for buoyancy to be
visible, \cite{Rev2,Vib}. In the unfluidized regime, effects like inertia, void filling models (true for vertical shaking) and convection are more important. Another difference to be noted is that buoyancy is not just a phenomenon of the larger particle climbing
up to the top of the pile but refers to a specific dependence of height on the relative sizes and densities.\\
On the other hand, the rising of larger objects in a sea of smaller grains due to the Brazil nut effect, \cite{BNE,Rev1,Rev2}, arises in several forms and many competing mechanisms influence the
motion of the larger particles within a bath (shape, size, forces between particles, shaking amplitude and direction, interstitial air and humidity). In \cite{Leb1,Leb2} a similar model to ours was used to investigate the Brazil nut effect. Sometimes segregation is a result of entropic forces which are strong in a
gravity free regime and when the frictional forces among bath particles and between bath and intruder are such that the entropy of a segregated state is higher than a highly mixed state, \cite{Granpatt}.

\begin{figure}[ht!]
\begin{center}
\includegraphics[height=8cm,angle=-90]{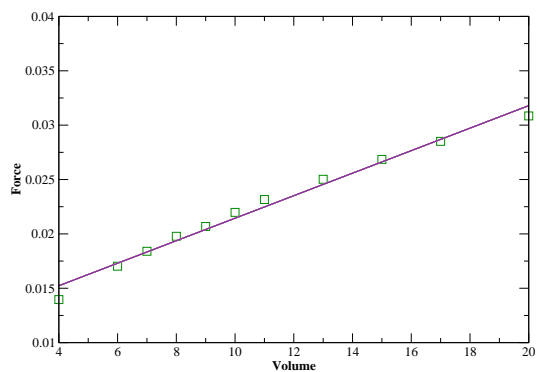}%,width=1in
\caption{Buoyancy force {\it vs } Rod length  $N$ at $\gamma = 15$, at height $y = 180$ and mass of bath particles
$m = 0.02$, $kT$ and $g$ taken to be unity}
\end{center}
\end{figure}

\subsection{Simulation results}\label{rema1}
Since the one-dimensional reduction of the full lattice fluid model has undergone some further simplifications, we have tested numerically whether our approximations appear reasonable.  In other words, we have compared the trajectories of the rod in our approximations with those of the true model.
The simulation was run on a chain of 300 sites and gravity $g$ and $kT$ are taken as unity. In this way we could also numerically verify Archimedes' law in the large $\gamma$ limit. In the fluid limit the buoyancy force varies linearly with the size $N$ of the rod, see Fig.~4. The straight line indicates that for such a large $\gamma$ the monomer bath is fluid-like. For a given height $y$ buoyancy force is calculated by estimating the weight of the rod which would exactly balance the force from the bath at that height.
In the fluid limit at equilibrium then, the weight of the rod is equal to the upward force(called buoyancy).

Before the fluid limit is reached buoyancy force varies with $\gamma$; for large $\gamma$ the buoyancy force tends to a steady value as given by Archimedes' law, see Fig.~5. That must be compared with Fig.3 in \cite{ex1}. The buoyancy force grows with $\gamma$ and after a certain critical value which in the simulation was $\gamma = 4.0 $, it saturates.
Fig.~5 shows the Archimedes' force \eqref{arc} corrected with the friction term as it acts in \eqref{lange1}, with the $2/(2+\gamma)$ kind of variation.

\subsection{Longer memory}\label{longmem}
Instead of considering memory only until one time step before, one could also take two, three or more time steps long memory. That means, to consider again \eqref{cex} and to take into account contributions from alternative histories. These contributions are all of smaller order, with each correction falling as an inverse power of $\gamma$. The power arises from performing the integral like in \eqref{fri} but now the time is a sum of exponential variables, so that we get corrections like  $\mu(\gamma)^n$.

\begin{figure}[ht!]
 \begin{center}
\includegraphics[height=8cm,angle=-90]{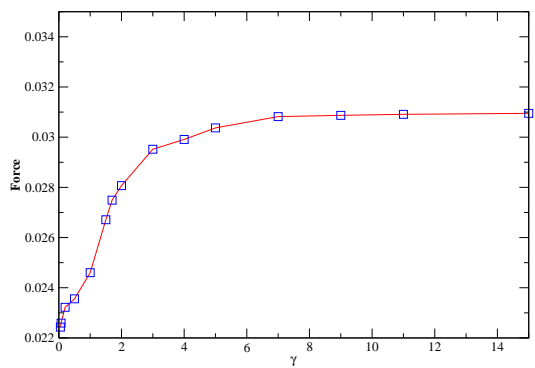}%,width=5in

%\begin{right}
\caption{Buoyancy force {\it vs} $\gamma$ for $N = 20$
at height $y = 180$ and mass of bath particles
$m = 0.02$, $kT$ and $g$ taken to be unity}
%\end{right}

\end{center}
\end{figure}

\subsection{Nonequilibrium seas} \label{seas}
Another type of correction to Archimedes' law comes from the possible nonequilibrium nature of the medium, even in the fluid limit.  Indeed, an essential aspect of our analysis above was that the sea of grains (monomers) reach their equilibrium between two moves of the rod (in the fluid limit), and that equilibrium is the same with or without the rod.  In other words, the stationary distribution of the constrained dynamics of the monomers given the rod's position gives exactly the same as conditioning the stationary distribution of the joint monomer--rod dynamics on the position of the rod.  That is only valid under the condition of detailed balance, see e.g.
Lemma 3.1 in \cite{hydro}.  In the present paper detailed balance is forced by the specific choice of density profile \eqref{dens} for which holds that
\begin{equation}\label{dba}
p\,d(y)\,(1-d(y+1)) = q\,d(y+1)\,(1-d(y))
\end{equation}
In the small mesh (continuum) limit this detailed balance condition \eqref{dba} becomes $d'(y) = -mg d(y)\,(1-d(y))/(kT)$ as repeatedly used in the derivation of the Langevin equations \eqref{lange} and \eqref{lange1}, and in the validity of the corresponding Einstein relation.
When detailed balance is violated and the granular sea shows an irreversible steady behavior, the motion remains much less understood.

\subsection{Collective effects}
A final source of corrections to Archimedes' law is due to the interaction with other intruders (rods).  Here we are really speaking about a whole new range of phenomena in which pairing of particles, \cite{monas} and more general collective effects as flocking \cite{nat}, can occur. The main underlying reason however is already visible from the simple analysis of the present paper.
The excluded volume effect of one rod not only creates a bias for itself to return to the place it was before (creating drag, \cite{dean}), but also creates space where another intruder can hop into,
and thus ``attracts'' other rods and intruders.  This granular-hydrodynamic interaction is long range and is expected to be proportional to the speeds of the rods, but more exploration is needed.
This interaction qualitatively resembles the long range hydrodynamic interaction between colloidal particles in suspensions. These interactions come through due to the Stokes-like force applied by the suspension on a moving colloid
which results in long-range interaction between two colloids connected through the Oseen tensor. Here, the collective behavior of multiple intruders results from simple exclusion and memory tracks left on the bath.

\section{Conclusions}

There are few ``mesoscopic'' toy-models of granular effects, and there is a wealth of experimental data.
We have presented a  theoretical model comprised of a large bath of small particles and the motion of a large body through this bath. We only consider positional degrees of freedom with a dissipative dynamics. Fluidization is controlled by the degree $\gamma$ of horizontal shaking; it modifies the relaxation time of the bath degrees of freedom.  The fast and the slow degrees of freedom get separated and the extended body undergoes buoyancy according to Archimedes' law.  Corrections before the fluid limit can be modeled in terms of memory effects, where the big particle is biased to fall back in the hole it left behind. This creates friction proportional to the velocity with a coefficient that is inversely proportional to the shaking amplitude\\
Granular medium is fertile ground for study of origin and behavior of statistical forces which arise due to coarse-graining. Two of those are studied here,
buoyancy due to the pressure gradients and additional memory effects creating friction. As an outlook, the study of effective forces and interactions out-of-equilibrium remains very much open.\\

\noindent {\bf Acknowledgments}:  We are grateful to Thijs Becker for useful discussions.


\begin{thebibliography}{99}


\small

\bibitem{ex1}
Huerta D.A., Sosa V., Vargas M.C., Ruiz-Su\'{a}rez J.C., (2005) Archimedes' principle
in fluidized granular systems,
{\sl Phys. Rev. E} {\bf 72}, 031307(1-5).

\bibitem{ex2}
Nichol K., Zanin A., Bastien R., Wandersman E., Hecke M.,
 (2009) Flow-induced Agitations create a Granular Fluid,
{\sl Phys. Rev. Lett.} {\bf 104}, 078302(1-4).

\bibitem{sim1} Shishodia N., Wassgren C.R.,
 (2001) Particle Segregation in Vibrofluidized Beds Due to Buoyant Forces,
{\sl Phys. Rev. Lett.} {\bf 87}, 084302(1-4).

\bibitem{phen}
Alam M., Trujillo L., Herrmann H.J., (2006) Hydrodynamic Theory for Reverse Brazil Nut Segregation and the Non-monotonic Ascension Dynamics,
{\sl J. Stat Phys.} {\bf 124}, 587--623.


\bibitem{hydro}
Ferrari P.A., Maes C., Ramos L., Redig F., (2000) On the Hydrodynamic equilibrium
 of a rod in a lattice fluid,
{\sl J. Phys. A: Math. Gen.} {\bf 33}, 4725--4740.

%\bibitem{mem}
%Josserand C., Tkachenko A.V., Mueth D.M., Jaeger H.M., (2000) Memory Effects in Granular Materials,
%{\sl Phys. Rev. Lett.} {\bf 85}, 3632--3635.

\bibitem{Mob}
Zik O., Stavans J., Rabin Y., (1992) Mobility of a Sphere in Vibrated Granular Media, 
{\sl Europhys. Letts.} {\bf 17(4)}, 315--319.

\bibitem{creep}
Candelier R., Dauchot O., (2009) Creep Motion of an Intruder within a Granular Glass Close to Jamming,
{\sl Phys. Rev. Lett. } {\bf 103}, 128001-(1-4).

\bibitem{slowdrag}
Geng J., Behringer R.P., (2005) Slow Drag in two-dimensional granular media,
{\sl Phys. Rev. E} {\bf 71}, 011302-(1-19).

\bibitem{Archi}
Archimedes, {\it On Floating Bodies I}, section 5,
{\it Works
of Archimedes} edited by \textit{Heath T.L.}, Cambridge:Cambridge University Press (1897).

\bibitem{Vib}
Huerta D.A., Ruiz-Su\'{a}rez J.C.,(2004) Vibration-induced granular segregation: a phenomenon driven by three mechanisms
{\sl Phys. Rev. Lett.} {\bf 92}, 114301.

\bibitem{Rev2}
Kudrolli A.,(2004) Size separation in vibrated granular matter,
{\sl Rep. Prog. Phys.} {\bf 67}, 209--248.

\bibitem{BNE}
Rosato A., Strandburg K.J., Prinz F., Swendsen R.P.,
 (1987) Why the Brazil Nuts are on Top: Size Segregation of Particulate Matter by Shaking,
{\sl Phys. Rev. Lett.} {\bf 58}, 1038--1040.

\bibitem{Rev1} Jaeger H.M, Nagel S.R., Behringer R.P.,
 (1996) Granular Solids, Liquids and Gases,
{\sl Reviews of Modern Physics} {\bf 68(4)}, 1259--1273.

\bibitem{Leb1} Alexander F.J., Lebowitz J.L., (1990) Driven diffusive
systems with a moving obstacle: a variation on the Brazil nuts problem,
{\sl J. Phys. A} {\bf 23}, L375--L381.

\bibitem{Leb2}
Alexander F.J., Lebowitz J.L., (1994) On the drift and
diffusion of a rod in a lattice fluid, {\sl J. Phys. A} {\bf 27}, 683--696.

\bibitem{Granpatt} Aranson I.S., Tsimring L.S.,
 (2009) Granular Patterns,
{\sl Oxford University Press}.

\bibitem{monas}
Mej\'{\i}a--Monasterio C., Oshanin G., (2011), Bias-- and bath--mediated pairing of particles driven through a quiescent medium, {\sl Soft Matter} {\bf 7}, 993-1000.

\bibitem{nat}
Pacheco-V\'{a}azquez F., Ruiz-Su\'{a}rez J.C., (2010) Cooperative dynamics in the penetration of a group of intruders in a granular medium,
Nature Communications {\bf 1}, Article number 123.

\bibitem{dean}
D\'emery V., Dean D.S., (2010) Drag forces on inclusions in classical fields with dissipative dynamics, {\it Eur.Phys. J.} {\bf 32}, 377--390.
















\end{thebibliography}
\end{document}